\documentclass[prd,aps,twocolumn,preprintnumbers, showpacs, nofootinbib,superscriptaddress,notitlepage]{revtex4-1}
\usepackage{amssymb,amsthm,amsmath}
\usepackage{textcomp}
\usepackage{subfig,graphicx}   
\usepackage{color}      
\usepackage{slashed}    
\usepackage{verbatim}
\usepackage[normalem]{ulem}
\usepackage{rotating}   
\usepackage{multirow}   
\begin{document}
\title{Global analysis of measured and unmeasured hadronic \\ two-body weak decays of antitriplet charmed baryons}
\author{  Zhi-Peng Xing}\email{Email:zpxing@nnu.edu.cn}
\affiliation{Department of Physics and Institute of Theoretical Physics, Nanjing Normal University, Nanjing, Jiangsu 210023, China}
\affiliation{Tsung-Dao Lee Institute,
Shanghai Jiao Tong University, Shanghai 200240, China}
\author{  Xiao-Gang He}\email{Email:hexg@sjtu.edu.cn}
\affiliation{Tsung-Dao Lee Institute,
Shanghai Jiao Tong University, Shanghai 200240, China}
\affiliation{School of Physics and Astronomy,
Shanghai Jiao Tong University, Shanghai 200240, China}\affiliation{National Center for Theoretical Sciences, Department of Physics, National Taiwan University, Taipei 10617, Taiwan}
\author{  Fei Huang}\email{Email:fhuang@sjtu.edu.cn}
\affiliation{School of Physics and Astronomy,
Shanghai Jiao Tong University, Shanghai 200240, China}
\author{ Chang Yang}\email{Email:15201868391@sjtu.edu.cn}
\affiliation{Tsung-Dao Lee Institute,
Shanghai Jiao Tong University, Shanghai 200240, China}

\begin{abstract}
A large amount of data on hadronic two body weak decays of anti-triplet charmed baryons $T_{c\bar 3}$ to an octet baryon $T_8$ and an octet or singlet pseudoscalar meson $P$, $T_{c \bar 3} \to T_8 P$, have been measured. The SU(3) flavor symmetry has been applied to study these decays to obtain insights about weak interactions for charm physics. However not all such decays needed to determine the SU(3) irreducible amplitudes have been measured forbidding a complete global analysis. Previously, it has been shown that data from measured decays can be used to do a global fit to determine all except one parity violating and one parity conserving amplitudes of the relevant SU(3) irreducible amplitudes causing 8 hadronic two body weak decay channels involving $\Xi^0_c$ to $\eta$ or $\eta'$ transitions undetermined. 
It is important to obtain information about these decays in order to guide experimental searches.  In this work using newly measured decay modes by BESIII and Belle in 2022, we carry out a global analysis and parameterize the unknown amplitudes to provide the ranges for the branching ratios of the 8 undetermined decays. Our results indicate that the SU(3) flavor symmetry can explain the measured data exceptionally well, with a remarkable minimal $\chi^2/d.o.f.$ of 1.21 and predict  80 observables in 45 decays for future experimental data to test.  We then vary the unknown SU(3) amplitudes to obtain the allowed range of branching ratios for the 8 undetermined decays. We find that some of them are within reach of near future experimental capabilities.  We urge our experimental colleagues to carry out related searches.
\end{abstract}

\maketitle

\section{Introduction}
Heavy flavor physics is an important part of particle physics study. Many interesting phenomena have been observed in recent years~\cite{Belle:2019oag,Belle:2019rba,LHCb:2021trn,LHCb:2022piu,Choudhury:2022ktg,LHCb:2023zxo,LHCb:2023cjr}. 
The decay of anti-triplet charmed baryons $T_{c\bar 3}$ to an octet baryon $T_8$ and an octet or singlet pseudoscalar meson $P$, denoted as $T_{c \bar 3} \to T_8 P$, has garnered attention from both theorists~\cite{Zhao:2018zcb,Zou:2019kzq,Jia:2019zxi,Geng:2019xbo,Zhao:2018mov,Xu:2021mkg,Huang:2021aqu,Zhao:2021sje,Groote:2021pxt,Li:2021iwf,Hsiao:2021nsc,Zhong:2022exp,Zhao:2023yuk,Liu:2023pyk,Wang:2023don} and experimentalists~\cite{BESIII:2017fim,BESIII:2018cvs,BESIII:2018cdl,Belle:2021crz,Belle:2021mvw,Li:2021bff,Lyu:2021biq}.
Last year,  several anti-triplet charmed baryons hadronic two body weak decays have been measured for the first time and several other decays were updated by BESIII~\cite{BESIII:2022izy,BESIII:2022tnm,BESIII:2022wxj,BESIII:2022bkj} and Belle~\cite{Belle:2021vyq,Belle:2022uod,Belle:2022bsi}. These update data are shown in Table.~\ref{table3}.

The new experimental data provide more information for a better understanding of the weak interactions of charmed particles. Due to the relatively low energies involved in charmed particle decays, perturbative QCD calculations of hadronic matrix elements become unreliable. Ways to address non-perturbative effects need to be found. Lattice QCD calculations hold promise for finally solving this problem, but a complete implementation is still premature at present.

Symmetry considerations may also provide some understanding of charmed baryon decays. A commonly used useful approach is the SU(3) flavor symmetry. With this approach, although it is not possible to calculate the absolute values of the decay amplitudes, it is possible to obtain relations between different decay amplitudes.  With a smaller number of amplitudes, when combined with experimental data, the amplitudes can be over constrained and predictions can be made to further test the approach. Of course, one needs to ensure the applicability of SU(3) flavor symmetry. Without a detailed understanding of the dynamics, one way to assess its validity is to see how well known data can be explained when the symmetry  is mposed, as measured by the quality of the fit through $\chi^2$ for each degree of freedom. In 2021, a global analysis of the SU(3) symmetry for anti-triplet charmed baryon hadronic two-body weak decays was carried out~\cite{Huang:2021aqu}. The analysis showed that the SU(3) symmetry can well explain the data and predict 87 observables in 49 decays, using only 16 input form factors.  Since then, many works~\cite{Hsiao:2021nsc,Zhong:2022exp} have also carried out global analyses for these decays.

Although experimental data are abundant, a complete SU(3) analysis for anti-triplet charmed baryon hadronic two-body decays is still not feasible. In Ref.\cite{Huang:2021aqu}, it was shown that  one pair of the SU(3) irreducible amplitude, the parity violating , and parity conserving amplitudes, correspond to $a^\prime$ defined in the paper is not constrained by measured data leading to  8 decays undetermined. They are $\Xi_c^0\to\Xi^0\eta^{(\prime)},\Xi_c^0\to\Sigma^0\eta^{(\prime)},\Xi_c^0\to\Lambda^0\eta^{(\prime)},\Xi_c^0\to n\eta^{(\prime)}$. Recently one of us (Yang) and his collaborator tried to calculate directly~\cite{Liu:2023pyk} for $\Xi_c^0\to\Sigma^0\eta^{(\prime)}$ processes. Unfortunately, the tree-level diagram they considered does not contribute to the undetermined parameter $a^\prime$ according to the topological diagram amplitude analysis~\cite{He:2018joe}. 

In this work, we will carry out an analysis by treating the amplitudes in $a^\prime$ as free parameters and letting them vary within a given range to predict the 8 unknown branching ratios. We will use the most recent data for global analysis to determine the other decay amplitudes. It is interesting to find that our new analysis confirms our previous result that SU(3) describes anti-triplet charmed baryon hadronic two body decays well with a $\chi^2/d.o.f.$ of 1.21. Within a reasonable range for $a^\prime$, some of the predicted branching ratios for these 8 unmeasured decays can be detected by near-future experiments.

\section{The SU(3) irreduciable amplitudes}\label{se_su3}

In the framework of SU(3) symmetry, the initial and final states of the charmed baryon decays we study can be expressed as SU(3) irreducible representations. Induced by the $c\to s/d$ due to W exchange, anti-triplet charmed baryons can decay into octet baryons and an octet or a singlet pseudoscalar meson. Specifically, in our work, we represent the anti-triplet charmed baryon $T_{c\bar 3}$, the octet baryon $T_8$ and the octet plus singlet pseudoscalar meson $P$ as
\begin{eqnarray}\
T_{c\bar3}&=&
\begin{pmatrix}
0& \Lambda_c^+ &\Xi_c^+ \\
-\Lambda_c^+ & 0&\Xi_c^0\\
-\Xi_c^+& -\Xi_c^0&0
\end{pmatrix},
P=
\begin{pmatrix}
\frac{\pi^0+\eta_q}{\sqrt{2}}& \pi^+ &K^+ \\
\pi^- & \frac{-\pi^0+\eta_q}{\sqrt{2}}&K^0\\
K^-& \bar{K}^0&\eta_s
\end{pmatrix},\notag\\
T_8&=&
\begin{pmatrix}
\frac{\Sigma^0}{\sqrt{2}}+\frac{\Lambda^0}{\sqrt{6}}& \Sigma^+ &p \\
\Sigma^- & -\frac{\Sigma^0}{\sqrt{2}}+\frac{\Lambda^0}{\sqrt{6}}&n\\
\Xi^-& \Xi^0&-\frac{2\Lambda^0}{\sqrt{6}}
\end{pmatrix}.
\end{eqnarray}
Here the anti-triplet charmed baryon can also be expressed as $T_{c\bar3}=(\Xi_c^0,-\Xi_c^+,\Lambda^+_c)$ and the $\eta_s$ and $\eta_q$ are the mixture of $\eta_1$ and $\eta_8$:
\begin{eqnarray}
\eta_8=\frac{1}{\sqrt{3}}\eta_q-\sqrt{\frac{2}{3}}\eta_s,\quad \eta_1=\sqrt{\frac{2}{3}}\eta_q+\frac{1}{\sqrt{3}}\eta_s.
\end{eqnarray}
The physical states $\eta$ and $\eta'$ are the mixture of the $\eta_q$ and $\eta_s$ as
\begin{eqnarray}
&&\begin{pmatrix}
\eta \\
\eta^\prime\\
\end{pmatrix}=\begin{pmatrix}
\cos\phi&-\sin\phi \\
\sin\phi&\cos\phi\\
\end{pmatrix}\begin{pmatrix}
\eta_q \\
\eta_s\\
\end{pmatrix},
\end{eqnarray}
with $\phi=(39.3\pm1.0)^{\circ}$~\cite{Gan:2020aco}.

The decays are induced by an effective weak interaction Hamiltonian composed of the charm quark, which is a singlet in SU(3), and the three light quarks, u, d and s, which form a triplet in SU(3)~\cite{Huang:2021aqu}. In the irreducible representation amplitude (IRA) method, the Hamiltonian composed of the three light quarks is decomposed into various irreducible representations, namely $3\otimes \bar 3\otimes 3=3\oplus 3\oplus\bar 6\oplus 15$.  In Ref. \cite{He:2018joe}, it was shown that the $\bar 3$ representation is extremely suppressed by the CKM matrix element and can be neglected. This is because only singly Cabibbo suppressed decays induced by $c\to u\bar d d$ and $c\to u\bar s s$ can contribute to $H_{\bar 3}$.  Since $c\to u\bar d d$ and $c\to u\bar s s$ transitions are approximately equal but with opposite signs, their sum is $V_{cd}V^*_{ud} + V_{cs}V_{us}^*= -V_{cb}V_{ub}^*$ , which leads to a very small coefficient for the $\bar 3$ component of the effective Hamiltonian that can be safely ignored. Consequently, only  $H_{\bar 6}$ and $H_{15}$ dominate the contribution to the Hamiltonian. The non-zero entries of these irreducible representations can be normalized as
\begin{eqnarray}
&&(H_{\bar 6})^{31}_2=-(H_{\bar 6})^{13}_2=1,\quad (H_{15})^{31}_2=(H_{15})^{13}_2=1,\notag\\
&& (H_{15})^{31}_3=(H_{15})^{13}_3=-(H_{15})^{21}_2=-(H_{15})^{12}_2=\sin\theta,\notag\\
&&(H_{\bar 6})^{31}_3=-(H_{\bar 6})^{13}_3=(H_{\bar 6})^{12}_2=-(H_{\bar 6})^{21}_2=\sin\theta,\notag\\
&&(H_{\bar 6})^{21}_3=-(H_{\bar 6})^{12}_3=(H_{15})^{21}_3=(H_{15})^{12}_3=\sin^2\theta,
\end{eqnarray}
with the assumption: $V_{ud}\approx V_{cs} \approx 1$ and $ V_{cd}\approx -V_{us}\approx -\sin\theta \approx 0.2265$.
Using the anti-symmetric tensor $\epsilon_{kmi}$ to contract $(H_{\bar 6})^{km}_j$ and $(T_{c\bar 3})^{[ij]}$,  we can define
\begin{eqnarray}\
(H_{\bar 6})_{\{ij\}}&=&\frac{1}{2}\epsilon_{kmi}(H_{\bar 6})^{km}_j=
\begin{pmatrix}
0&0&0 \\
0& 1& \sin\theta\\
0& \sin\theta&-\sin^2\theta
\end{pmatrix},\notag\\
 (T_{c\bar 3})_i&=&\epsilon_{ijk}(T_{c\bar 3})^{[jk]}.
\end{eqnarray}
Following the same method and analysis in ref.\cite{Huang:2021aqu}, the SU(3) invariant decay amplitudes can be written as
 \begin{eqnarray}
\mathcal{M}=&&a_{15} \times(T_{c\bar{3}})_i(H_{\overline{15}})^{\{ik\}}_j(\overline{T_8})^j_kP^l_l\notag\\
&+&b_{15} \times(T_{c\bar{3}})_i(H_{\overline{15}})^{\{ik\}}_j(\overline{T_8})^l_kP^j_l\notag\\
&+&c_{15} \times(T_{c\bar{3}})_i(H_{\overline{15}})^{\{ik\}}_j(\overline{T_8})^j_lP^l_k\notag\\
&+&d_{15} \times(T_{c\bar{3}})_i(H_{\overline{15}})^{\{jk\}}_l(\overline{T_8})^l_jP^i_k\notag\\
&+&e_{15} \times(T_{c\bar{3}})_i(H_{\overline{15}})^{\{jk\}}_l(\overline{T_8})^i_jP^l_k\notag\\
&+&a_{6} \times(T_{c\bar{3}})^{[ik]}(H_{\overline{6}})_{\{ij\}}
(\overline{T_8})^j_kP^l_l\notag\\
&+&b_{6} \times(T_{c\bar{3}})^{[ik]}(H_{\overline{6}})_{\{ij\}}(\overline{T_8})^l_kP^j_l\notag\\
&+&c_{6} \times(T_{c\bar{3}})^{[kl]}(H_{\overline{6}})_{\{ij\}}(\overline{T_8})^j_lP^l_k\notag\\&+&d_{6} \times(T_{c\bar{3}})^{[ik]}(H_{\overline{6}})_{\{ij\}}(\overline{T_8})^i_kP^j_l.\label{su3}
\end{eqnarray}

The specific process is derived by expanding these amplitudes into a linear combination of SU(3) irreducible amplitudes. These SU(3) amplitudes, which are given in  Ref.~\cite{Huang:2021aqu} and Table.~\ref{table1} and Table.~\ref{tableta}.  In Table.~\ref{tableta}, we give the SU(3) amplitudes of anti-triplet charmed baryon decays into an octet baryon and  $\eta$($\eta^\prime$). One can see that only decays with $\eta$($\eta^\prime$) in the final states depend on the parameters $a_6$ and $a_{15}$. 

\section{The global fit analysis}\label{se_gf}

After performing the global fit presented in Ref.~\cite{Huang:2021aqu}, a significant amount of experimental data has been updated, and some previously unmeasured decay channels are now available experimentally which are shown in Table \ref{table3}. In addition, several theoretical studies have focused on related decays~\cite{Hsiao:2021nsc,Zhong:2022exp}. As a result of these advancements, it is necessary to address the gaps in the previous analysis and update it so as to obtain more insights from the newly available data.

\begin{widetext}
\begin{table*}[htbp!]
\caption{Experimental data and fitting results of anti-triplet charmed baryons two-body decays.  The fitting results for the parity violating $f^q_i$ and parity conserving $g^q_i$ form factors are also shown in the last five rows.}\label{table3}
\begin{tabular}{|c|c|c|c|c|c|c|c|c|c|}\hline\hline
\multirow{2}{*}{Channel} &\multicolumn{4}{c|}{ Branching ratio}\cr\cline{2-5}
&Latest measurement in 2022($\%$)&Experimental data($\%$)& Previous work($\%$) \cite{Huang:2021aqu}& This work($\%$)\\\hline
$\Lambda^{+}_{c}\to p K_S^0 $ &$-$& $1.59\pm0.08$ \cite{ParticleDataGroup:2022pth}&$1.587\pm0.077$ &$1.606\pm0.077$\\\hline
$\Lambda^{+}_{c}\to p\eta $ &$-$&$0.142\pm0.012$\cite{ParticleDataGroup:2022pth}&$0.127\pm0.024$&$0.141\pm0.011$\\\hline
\multirow{2}{*}{$\Lambda^{+}_{c}\to p\eta^\prime $} &$0.0562^{+0.0246}_{-0.0204}\pm0.0026$\cite{BESIII:2022izy}&\multirow{2}{*}{$0.0484\pm0.0091$\cite{Belle:2021vyq,BESIII:2022izy}}&\multirow{2}{*}{$0.27\pm0.38$}&\multirow{2}{*}{$0.0468\pm0.0066$}\cr\cline{2-2}
&$0.0473\pm0.0082\pm0.0046\pm0.0024$\cite{Belle:2021vyq}&&&\\\hline
$\Lambda^{+}_{c}\to \Lambda \pi^+ $ &$1.31\pm0.08\pm0.05$\cite{BESIII:2022bkj}& $1.30\pm0.06$\cite{ParticleDataGroup:2022pth,BESIII:2022bkj}&$1.307\pm0.069$&$1.328\pm0.055$\\\hline
$\Lambda^{+}_{c}\to \Sigma^0\pi^+ $ &$1.22\pm0.08\pm0.07$\cite{BESIII:2022bkj}& $1.27\pm0.06$\cite{ParticleDataGroup:2022pth,BESIII:2022bkj}&$1.272\pm0.056$&$1.260\pm0.046$\\\hline
$\Lambda^{+}_{c}\to \Sigma^{+}\pi^0 $ &$-$& $1.25\pm0.10$\cite{ParticleDataGroup:2022pth}&$1.283\pm0.057$&$1.274\pm0.047$\\\hline
$\Lambda^{+}_{c}\to \Xi^{0}K^{+} $ &$-$& $0.55\pm0.07$\cite{ParticleDataGroup:2022pth}&$0.548\pm0.068$&$0.430\pm0.030$\\\hline
\multirow{2}{*}{$\Lambda^{+}_{c}\to \Lambda^{0}K^{+} $} &$0.0621\pm0.0044\pm0.0026\pm0.0034$\cite{BESIII:2022tnm}& \multirow{2}{*}{$0.064\pm0.003$\cite{ParticleDataGroup:2022pth,BESIII:2022tnm,Belle:2022uod}}&\multirow{2}{*}{$0.064\pm0.010$}&\multirow{2}{*}{$0.0646\pm0.0028$}\cr\cline{2-2}
&$0.0657\pm0.0017\pm0.0011\pm0.0035$\cite{Belle:2022uod}&&&\\\hline
$\Lambda^{+}_{c}\to \Sigma^{+}\eta $ &$0.416\pm0.075\pm0.021\pm0.033$\cite{Belle:2022bsi}& $0.32\pm0.043$\cite{ParticleDataGroup:2022pth,Belle:2022bsi}&$0.45\pm0.19$&$0.329\pm0.042$\\\hline
$\Lambda^{+}_{c}\to \Sigma^{+}\eta^\prime $ &$0.314\pm0.035\pm0.011\pm0.025$\cite{Belle:2022bsi}& $0.437\pm0.084$\cite{ParticleDataGroup:2022pth,Belle:2022bsi}&$1.5\pm0.6$&$0.444\pm0.070$\\\hline
\multirow{2}{*}{$\Lambda^{+}_{c}\to \Sigma^{0}K^+ $ }&$0.047\pm0.009\pm0.001\pm0.003$\cite{BESIII:2022wxj}& \multirow{2}{*}{$0.0382\pm0.0025$\cite{ParticleDataGroup:2022pth,BESIII:2022wxj,Belle:2022uod}}&\multirow{2}{*}{$0.0504\pm0.0056$}&\multirow{2}{*}{$0.0381\pm0.0017$}\cr\cline{2-2}
&$0.0358\pm0.0019\pm0.0006\pm0.0019$\cite{Belle:2022uod}&&&\\\hline
$\Lambda^+_c\to n\pi^+$ &$0.066\pm0.012\pm0.004$\cite{BESIII:2022bkj}& $0.066\pm0.0126$\cite{BESIII:2022bkj}&$0.035\pm0.011$&$0.0651\pm0.0026$\\\hline
$\Lambda^+_c\to \Sigma^+K_s^0$ &$0.048\pm0.014\pm0.002\pm0.003$\cite{BESIII:2022wxj}& $0.048\pm0.0145$\cite{BESIII:2022wxj}&$0.0103\pm0.0042$&$0.0327\pm0.0029$\\\hline
$\Xi^{+}_{c}\to \Xi^{0}\pi^+ $ &$-$& $1.6\pm0.8$\cite{ParticleDataGroup:2022pth}&$0.54\pm0.18$&$0.887\pm0.080$\\\hline
$\Xi^{0}_{c}\to \Lambda K_S^0 $ &$-$& $0.32\pm0.07$\cite{ParticleDataGroup:2022pth}&$0.334\pm0.065$&$0.261\pm0.043$\\\hline
$\Xi^{0}_{c}\to \Xi^- \pi^+ $ &$-$& $1.43\pm0.32$\cite{ParticleDataGroup:2022pth}&$1.21\pm0.21$&$1.06\pm0.20$\\\hline
$\Xi^{0}_{c}\to \Xi^- K^+ $ &$-$& $0.039\pm0.012$\cite{ParticleDataGroup:2022pth}&$0.047\pm0.0083$&$0.0474\pm0.0090$\\\hline
$\Xi_c^0\to\Sigma^0 K^0_S$ &$-$& $0.054\pm0.016$\cite{ParticleDataGroup:2022pth}&$0.069\pm0.024$&$0.054\pm0.016$\\\hline
$\Xi_c^0\to\Sigma^+ K^-$ &$-$& $0.18\pm0.04$\cite{ParticleDataGroup:2022pth}&$0.221\pm0.068$&$0.188\pm0.039$\\\hline\hline
\multirow{2}{*}{Channel} &\multicolumn{4}{c|}{ Asymmetry parameter $\alpha$}\cr\cline{2-5}
&Lastest measurement in 2022&Experimental data& Previous work\cite{Huang:2021aqu} & This work\\\hline
$\alpha(\Lambda^{+}_{c}\to p K_S^0)$&$-$&$0.18\pm0.45$\cite{ParticleDataGroup:2022pth}&$0.19\pm0.41$&$0.49\pm0.20$\\\hline
$\alpha(\Lambda^{+}_{c}\to \Lambda \pi^+)$&$-0.755\pm0.005\pm0.003$\cite{Belle:2022uod}&$-0.755\pm0.0058$\cite{ParticleDataGroup:2022pth,Belle:2022uod}&$-0.841\pm0.083$&$-0.7542\pm0.0058$\\\hline
$\alpha(\Lambda^{+}_{c}\to \Sigma^{0}\pi^+)$&$-0.463\pm0.016\pm0.008$\cite{Belle:2022uod}&$-0.466\pm0.0178$\cite{ParticleDataGroup:2022pth,Belle:2022uod}&$-0.605\pm0.088$&$-0.471\pm0.015$\\\hline
$\alpha(\Lambda^{+}_{c}\to  \Sigma^{+}\pi^0)$&$-0.48\pm0.02\pm0.02$\cite{Belle:2022bsi}&$-0.48\pm0.03$\cite{ParticleDataGroup:2022pth,Belle:2022bsi}&$-0.603\pm0.088$&$-0.468\pm0.015$\\\hline
$\alpha(\Xi^{0}_{c}\to \Xi^{-} \pi^{+})$&$-$&$-0.64\pm0.051$\cite{ParticleDataGroup:2022pth}&$-0.56\pm0.32$&$-0.654\pm0.050$\\\hline
$\alpha(\Lambda^{+}_{c}\to \Sigma^{0}K^+)$&$-0.54\pm0.18\pm0.09$\cite{Belle:2022uod}&$-0.54\pm0.20$\cite{Belle:2022uod}&$-0.953\pm0.040$&$-0.9958\pm0.0045$\\\hline
$\alpha(\Lambda^{+}_{c}\to \Lambda K^+)$&$-0.585\pm0.049\pm0.018$\cite{Belle:2022uod}&$-0.585\pm0.052$\cite{Belle:2022uod}&$-0.24\pm0.15$&$-0.545\pm0.046$\\\hline
$\alpha(\Lambda^{+}_{c}\to \Sigma^{+}\eta)$&$-0.99\pm0.03\pm0.05$\cite{Belle:2022bsi}&$-0.99\pm0.058$\cite{Belle:2022bsi}&$0.3\pm3.8$&$-0.970\pm0.046$\\\hline
$\alpha(\Lambda^{+}_{c}\to \Sigma^{+}\eta^\prime)$&$-0.46\pm0.06\pm0.03$\cite{Belle:2022bsi}&$-0.46\pm0.067$\cite{Belle:2022bsi}&$0.8\pm1.9$&$-0.455\pm0.064$\\\hline\hline
\multicolumn{5}{|c|}{ SU(3) symmetry parameters from fitting ($\chi^{2}$/d.o.f.=1.21)}\cr\hline
\multirow{2}{*}{Vector(f) }&$f^a=0.0155\pm0.0040$&$f^{b}_{6}=0.0215\pm0.0092$&$f^{c}_{6}= 0.0356\pm0.0071$&$f^d_6=-0.0138\pm0.0080$\cr\cline{2-5}
&$f^{b}_{15}=-0.0161\pm0.0042$&$f^{c}_{15}=0.0149\pm0.0080$&$f^d_{15}=-0.0253\pm0.0031$&$f^e_{15}= 0.0798\pm0.0087$\\\hline
\multirow{2}{*}{Axial-vector(g)}&$g^a=-0.039\pm0.012$&$g^{b}_{6}=-0.240\pm0.011$&$g^{c}_{6}= 0.121\pm0.019$&$g^d_6=-0.067\pm0.014$\cr\cline{2-5}
&$g^{b}_{15}=0.1134\pm0.0074$&$g^{c}_{15}=0.014\pm0.018$&$g^d_{15}=-0.0387\pm0.0085$&$g^e_{15}= 0.0209\pm0.0092$\\\hline
\end{tabular}
\end{table*}

\begin{table*}[htbp!]
\caption{SU(3) amplitudes and predicted branching fractions (the third column) and polarization parameters (the fourth column) of anti-triplet charmed baryons  decay into an octet baryon and an octet meson.}\label{table1}\begin{tabular}{|c|c|c|c|c|c|c|c}\hline\hline
Channel &SU(3) amplitude& Branching ratio($10^{-2}$)&$\alpha$\\\hline \hline
$\Lambda^{+}_{c}\to \Sigma^{0}  \pi^{+} $ & $ (-b_6+b_{15}+c_6-c_{15}+d_6)/\sqrt{2}$&$1.260\pm0.046$&$-0.470\pm0.015$\\\hline
$\Lambda^{+}_{c}\to \Lambda  \pi^{+} $ & $ -(b_6-b_{15}+c_6-c_{15}+d_6+2 e_{15})/\sqrt{6}$&$1.328\pm0.055$&$-0.7542\pm0.0058$\\\hline
$\Lambda^{+}_{c}\to \Sigma^{+}  \pi^{0} $ & $ (b_6-b_{15}-c_6+c_{15}-d_6)/\sqrt{2}$&$1.274\pm0.047$&$-0.468\pm0.015$\\\hline
$\Lambda^{+}_{c}\to p  K_{S}^{0} $ & $ ( \sin^2\theta \left(-d_6+d_{15}+e_{15}\right)+b_6-b_{15}-e_{15})/\sqrt{2}$&$1.606\pm0.077$&$0.49\pm0.20$\\\hline
$\Lambda^{+}_{c}\to \Xi^{0}  K^{+} $ & $ -c_6+c_{15}+d_{15}$&$0.430\pm0.030$&$0.955\pm0.018$\\\hline
$\Xi^{+}_{c}\to \Sigma^{+}  K_{S}^{0} $ & $  (\sin^2\theta \left(b_6-b_{15}-e_{15}\right)-d_6+d_{15}+e_{15})/\sqrt{2}$&$0.77\pm0.32$&$0.29\pm0.29$\\\hline
$\Xi^{+}_{c}\to \Xi^{0}  \pi^{+} $ & $ -d_6-d_{15}-e_{15}$&$0.887\pm0.080$&$-0.902\pm0.039$\\\hline
$\Xi^{0}_{c}\to \Sigma^{0}  K_{S}^{0} $ & $  (-\sin^2\theta \left(b_6+b_{15}-e_{15}\right)+(c_6+c_{15}+d_6-e_{15}))/2$&$0.054\pm0.016$&$-0.75\pm0.24$\\\hline
\multirow{2}{*}{$\Xi^{0}_{c}\to \Lambda  K^0_S$ }  &$ \sqrt{3}\sin^2\theta \left(b_6+b_{15}-2 c_6-2 c_{15}-2 d_6+e_{15}\right)/6$&\multirow{2}{*}{$0.261\pm0.043$}&\multirow{2}{*}{$0.984\pm0.084$}\cr
&$+ \sqrt{3}(2 b_6+2 b_{15}-c_6-c_{15}-d_6-e_{15})/6$&&\\\hline
$\Xi^{0}_{c}\to \Sigma^{+}  K^{-} $ & $ c_6+c_{15}+d_{15}$&$0.188\pm0.039$&$0.98\pm0.20$\\\hline
$\Xi^{0}_{c}\to \Xi^{-}  \pi^{+} $ & $ b_6+b_{15}+e_{15}$&$1.06\pm0.20$&$-0.654\pm0.050$\\\hline
$\Xi^{0}_{c}\to \Xi^{0}  \pi^{0} $ & $ (-b_6-b_{15}+d_6+d_{15})/\sqrt{2}$&$0.130\pm0.051$&$-0.28\pm0.18$\\\hline\hline
$\Lambda^{+}_{c}\to \Sigma^{0}  K^{+} $   & $ \sin\theta \left(-b_6+b_{15}+d_6+d_{15}\right)/\sqrt{2}$&$0.0381\pm0.0017$&$-0.9959\pm0.0044$\\\hline
$\Lambda^{+}_{c}\to \Lambda  K^{+} $  & $ -\sin\theta \left(b_6-b_{15}-2 c_6+2 c_{15}+d_6+3 d_{15}+2 e_{15}\right)/\sqrt{6}$&$0.0646\pm0.0028$&$-0.545\pm0.046$\\\hline
$\Lambda^{+}_{c}\to \Sigma^{+} K^0_S/K^0_L $   & $ \sin\theta \left(-b_6+b_{15}+d_6-d_{15}\right)/\sqrt{2}$&$0.0327\pm0.0029$&$-0.52\pm0.11$\\\hline
$\Lambda^{+}_{c}\to p  \pi^{0} $          & $ \sin\theta \left(-c_6+c_{15}-d_6+e_{15}\right)/\sqrt{2}$&$0.021\pm0.010$&$-0.21\pm0.18$\\\hline
$\Lambda^{+}_{c}\to n  \pi^{+}$           & $ -\sin\theta \left(c_6-c_{15}+d_6+e_{15}\right)$&$0.0651\pm0.0026$&$0.533\pm0.047$\\\hline
$\Xi^{+}_{c}\to \Sigma^{0}  \pi^{+} $     & $ -\sin\theta \left(b_6-b_{15}-c_6+c_{15}+d_{15}+e_{15}\right)/\sqrt{2}$&$0.3194\pm0.0088$&$-0.728\pm0.018$\\\hline
$\Xi^{+}_{c}\to \Lambda  \pi^{+} $    & $ \sin\theta \left(-b_6+b_{15}-c_6+c_{15}+2 d_6+3 d_{15}+e_{15}\right)/\sqrt{6}$&$0.0222\pm0.0032$&$-0.16\pm0.17$\\\hline
$\Xi^{+}_{c}\to \Sigma^{+}  \pi^{0} $     & $ \sin\theta \left(b_6-b_{15}-c_6+c_{15}-d_{15}-e_{15}\right)/\sqrt{2}$&$0.247\pm0.020$&$0.46\pm0.19$\\\hline
$\Xi^{+}_{c}\to p  K^0_S/K^0_L $          & $ \sin\theta \left(-b_6+b_{15}+d_6-d_{15}\right)$&$0.177\pm0.016$&$-0.361\pm0.081$\\\hline
$\Xi^{+}_{c}\to \Xi^{0}  K^{+} $          & $ -\sin\theta \left(c_6-c_{15}+d_6+e_{15}\right)$&$0.1361\pm0.0063$&$0.371\pm0.036$\\\hline
$\Xi^{0}_{c}\to \Sigma^{0}  \pi^{0} $     & $ -\frac{1}{2} \sin\theta \left(b_6+b_{15}+c_6+c_{15}-d_{15}-e_{15}\right)$&$0.00014\pm0.00030$&$0.3\pm2.3$\\\hline
$\Xi^{0}_{c}\to \Lambda  \pi^{0} $    & $ \sin\theta \left(b_6+b_{15}+c_6+c_{15}-2 d_6-3 d_{15}+e_{15}\right)/2 \sqrt{3}$&$0.0375\pm0.0076$&$0.74\pm0.16$\\\hline
$\Xi^{0}_{c}\to \Sigma^{+}  \pi^{-} $     & $ -\sin\theta \left(c_6+c_{15}+d_{15}\right)$&$0.0116\pm0.0026$&$0.96\pm0.25$\\\hline
$\Xi^{0}_{c}\to p  K^{-} $                & $ \sin\theta \left(c_6+c_{15}+d_{15}\right)$&$0.0138\pm0.0045$&$0.89\pm0.38$\\\hline
$\Xi^{0}_{c}\to \Sigma^{-}  \pi^{+} $     & $ -\sin\theta \left(b_6+b_{15}+e_{15}\right)$&$0.057\pm0.011$&$-0.723\pm0.050$\\\hline
$\Xi^{0}_{c}\to n  K^0_S/K^0_L $& $ \sin\theta \left(-b_6-b_{15}+c_6+c_{15}+d_6\right)$&$0.0234\pm0.0060$&$0.66\pm0.34$\\\hline
$\Xi^{0}_{c}\to \Xi^{-}  K^{+} $          & $ \sin\theta \left(b_6+b_{15}+e_{15}\right)$&$0.0474\pm0.0090$&$-0.610\pm0.048$\\\hline
$\Xi^{0}_{c}\to \Xi^{0}  K^0_S/K^0_L $          & $ \sin\theta \left(b_6+b_{15}-c_6-c_{15}-d_6\right)$&$0.0114\pm0.0023$&$0.87\pm0.30$\\\hline\hline
$\Lambda^{+}_{c}\to p  K^0_L $ 		& $ (\sin^2\theta \left(-d_6+d_{15}+e_{15}\right)-b_6+b_{15}+e_{15})/\sqrt{2}$&$1.688\pm0.080$&$0.56\pm0.20$\\\hline
$\Lambda^{+}_{c}\to n  K^{+} $ 		& $ \sin^2\theta \left(d_6+d_{15}+e_{15}\right)$&$0.001022\pm0.000091$&$-0.980\pm0.019$\\\hline
$\Xi^{+}_{c}\to \Sigma^{0}  K^{+} $ & $ \sin^2\theta \left(b_6-b_{15}+e_{15}\right)/\sqrt{2}$&$0.01156\pm0.00033$&$-0.9961\pm0.0014$\\\hline
$\Xi^{+}_{c}\to \Lambda  K^{+} $& $\sin^2\theta \left(b_6-b_{15}-2 c_6+2 c_{15}-2 d_6-e_{15}\right)/\sqrt{6}$&$0.00441\pm0.00019$&$0.624\pm0.033$\\\hline
$\Xi^{+}_{c}\to \Sigma^{+}  K^0_L $ & $ (\sin^2\theta \left(b_6-b_{15}-e_{15}\right)+d_6-d_{15}-e_{15})/\sqrt{2}$&$0.95\pm0.35$&$0.57\pm0.28$\\\hline
$\Xi^{+}_{c}\to p  \pi^{0} $ 		& $\sin^2\theta \left(c_6-c_{15}+d_{15}\right)/\sqrt{2}$&$0.00046\pm0.00021$&$-0.29\pm0.38$\\\hline
$\Xi^{+}_{c}\to n  \pi^{+} $ 		& $ \sin^2\theta \left(c_6-c_{15}-d_{15}\right)$&$0.00619\pm0.00040$&$0.945\pm0.020$\\\hline
$\Xi^{0}_{c}\to \Sigma^{0}  K_{L}^{0}$ & $  (-\sin^2\theta \left(b_6+b_{15}-e_{15}\right)-(c_6+c_{15}+d_6-e_{15}))/2$&$0.069\pm0.019$&$-0.51\pm0.29$\\\hline
\multirow{2}{*}{$\Xi^{0}_{c}\to \Lambda  K^0_L$ }  &$ \sqrt{3}\sin^2\theta \left(b_6+b_{15}-2 c_6-2 c_{15}-2 d_6+e_{15}\right)/6$&\multirow{2}{*}{$0.243\pm0.043$}&\multirow{2}{*}{$0.996\pm0.043$}\cr
&$- \sqrt{3}(2 b_6+2 b_{15}-c_6-c_{15}-d_6-e_{15})/6$&&\\\hline
$\Xi^{0}_{c}\to p  \pi^{-} $ 		& $ \sin^2\theta \left(c_6+c_{15}+d_{15}\right)$&$0.00082\pm0.00029$&$0.87\pm0.40$\\\hline
$\Xi^{0}_{c}\to \Sigma^{-}  K^{+} $ & $ \sin^2\theta \left(b_6+b_{15}+e_{15}\right)$&$0.00258\pm0.00049$&$-0.689\pm0.050$\\\hline
$\Xi^{0}_{c}\to n  \pi^{0} $ 		& $ -\sin^2\theta \left(c_6+c_{15}-d_{15}\right)/\sqrt{2}$&$0.00194\pm0.00031$&$0.9997\pm0.0091$\\\hline
\hline
\end{tabular}
\end{table*}

\begin{table*}[htbp!]
\caption{SU(3) amplitudes and  predicted branching fractions (the third column) and polarization parameters (the fourth column) of anti-triplet charmed baryons  decay into an octet baryon and  $\eta$ or $\eta^\prime$. In this table ``$-$" represent the channel can not be predicted due to the limit of experimental data.}\label{tableta}
\begin{tabular}{|c|c|c|c|c|c|c|c}\hline\hline
Channel &SU(3) amplitude& Branching fraction($10^{-2}$)&$\alpha$\\\hline \hline
$\Lambda^{+}_{c}\to \Sigma^{+}  \eta $ & $\cos\phi(-2 a_6+2 a_{15}-b_6+b_{15}-c_6+c_{15}+d_6)/\sqrt{2}-\sin\phi( -a_6+a_{15}+d_{15})$&$0.329\pm0.042$&$-0.970\pm0.046$\\\hline
$\Lambda^{+}_{c}\to \Sigma^{+}  \eta^\prime $ & $\sin\phi(-2 a_6+2 a_{15}-b_6+b_{15}-c_6+c_{15}+d_6)/\sqrt{2}+\cos\phi( -a_6+a_{15}+d_{15})$&$0.444\pm0.070$&$-0.455\pm0.064$\\\hline
\multirow{2}{*}{$\Lambda^{+}_{c}\to p  \eta $ }       & $\sin\theta\big(\cos\phi \left(-2 a_6+2 a_{15}-c_6+c_{15}+d_6-e_{15}\right)/\sqrt{2}$&\multirow{2}{*}{$0.141\pm0.011$}&\multirow{2}{*}{$0.93\pm0.11$}\cr
&$- \sin\phi \left(-a_6+a_{15}-b_6+b_{15}+d_{15}+e_{15}\right)\big)$&&\\\hline
\multirow{2}{*}{$\Lambda^{+}_{c}\to p  \eta^\prime $}         & $\sin\theta\big(\sin\phi \left(-2 a_6+2 a_{15}-c_6+c_{15}+d_6-e_{15}\right)/\sqrt{2}$&\multirow{2}{*}{$0.0468\pm0.0066$}&\multirow{2}{*}{$-0.990\pm0.018$}\cr
&$+ \cos\phi \left(-a_6+a_{15}-b_6+b_{15}+d_{15}+e_{15}\right)\big)$&&\\\hline
\multirow{2}{*}{$\Xi^{+}_{c}\to \Sigma^{+} \eta $ }   & $\sin\theta \big(\cos\phi\left(-2 a_6+2 a_{15}-b_6+b_{15}-c_6+c_{15}+d_{15}+e_{15}\right)/\sqrt{2} $&\multirow{2}{*}{$0.114\pm0.022$}&\multirow{2}{*}{$0.97\pm0.11$}\cr
&$-\sin\phi \left(-a_6+a_{15}+d_6-e_{15}\right)\big)$&&\\\hline
\multirow{2}{*}{$\Xi^{+}_{c}\to \Sigma^{+}  \eta^\prime $ }   &$\sin\theta \big(\sin\phi\left(-2 a_6+2 a_{15}-b_6+b_{15}-c_6+c_{15}+d_{15}+e_{15}\right)/\sqrt{2} $&\multirow{2}{*}{$0.125\pm0.022$}&\multirow{2}{*}{$-0.456\pm0.070$}\cr
&$+\cos\phi \left(-a_6+a_{15}+d_6-e_{15}\right)\big)$&&\\\hline
\multirow{2}{*}{$\Xi^{+}_{c}\to p  \eta $} 		& $ \sin^2\theta \big(\cos\phi\left(2 a_6-2 a_{15}+c_6-c_{15}-d_{15}\right)/\sqrt{2}$&\multirow{2}{*}{$0.00938\pm0.00071$}&\multirow{2}{*}{$-0.003\pm0.61$}\cr
&$-\sin\phi\left(a_6-a_{15}+b_6-b_{15}-d_6\right)\big)$&&\\\hline
\multirow{2}{*}{$\Xi^{+}_{c}\to p  \eta^\prime $} 		& $ \sin^2\theta \big(\sin\phi\left(2 a_6-2 a_{15}+c_6-c_{15}-d_{15}\right)/\sqrt{2}$&\multirow{2}{*}{$0.0095\pm0.0011$}&\multirow{2}{*}{$-0.9981\pm0.0058$}\cr
&$+\cos\phi\left(a_6-a_{15}+b_6-b_{15}-d_6\right)\big)$&&\\\hline
$\Xi^{0}_{c}\to \Xi^{0}  \eta $ & $ \cos\phi(2 a_6+2 a_{15}+b_6+b_{15}-d_6+d_{15})/\sqrt{2}-\sin\phi(a_6+a_{15}+c_6+c_{15})$&-&-\\\hline
$\Xi^{0}_{c}\to \Xi^{0}  \eta^\prime $ & $ \sin\phi(2 a_6+2 a_{15}+b_6+b_{15}-d_6+d_{15})/\sqrt{2}+\cos\phi(a_6+a_{15}+c_6+c_{15})$&-&-\\\hline
\multirow{2}{*}{$\Xi^{0}_{c}\to \Sigma^{0}  \eta $}    & $ \sin\theta \big(\
cos\phi\left(2 a_6+2 a_{15}+b_6+b_{15}+c_6+c_{15}+d_{15}-e_{15}\right)/2$&\multirow{2}{*}{-}&\multirow{2}{*}{-}\cr
&$-\sin\phi \left(a_6+a_{15}-d_6+e_{15}\right)/\sqrt{2}\big)$&&\\\hline
\multirow{2}{*}{$\Xi^{0}_{c}\to \Sigma^{0}  \eta^\prime$}    & $ \sin\theta (\sin\phi\left(2 a_6+2 a_{15}+b_6+b_{15}+c_6+c_{15}+d_{15}-e_{15}\right)/2$&\multirow{2}{*}{-}&\multirow{2}{*}{-}\cr
&$-\cos\phi \left(a_6+a_{15}-d_6+e_{15}\right)/\sqrt{2}\big)$&&\\\hline
\multirow{2}{*}{$\Xi^{0}_{c}\to \Lambda  \eta $}   & $  \big( -\cos\phi \left(6 a_6+6 a_{15}+b_6+b_{15}+c_6+c_{15}-2 d_6+3 d_{15}+e_{15}\right)/(2 \sqrt{3})$&\multirow{2}{*}{-}&\multirow{2}{*}{-}\cr
&$-\sin\phi\left(-3 a_6-3 a_{15}-2 b_6-2 b_{15}-2 c_6-2 c_{15}+d_6+e_{15}\right)/\sqrt{6}\big)\sin\theta$&&\\\hline
\multirow{2}{*}{$\Xi^{0}_{c}\to \Lambda  \eta^\prime$ }  &$  \big( -\sin\phi \left(6 a_6+6 a_{15}+b_6+b_{15}+c_6+c_{15}-2 d_6+3 d_{15}+e_{15}\right)/(2 \sqrt{3})$&\multirow{2}{*}{-}&\multirow{2}{*}{-}\cr
&$+\cos\phi\left(-3 a_6-3 a_{15}-2 b_6-2 b_{15}-2 c_6-2 c_{15}+d_6+e_{15}\right)/\sqrt{6}\big)\sin\theta$&&\\\hline
\multirow{2}{*}{$\Xi^{0}_{c}\to n \eta $ }       & $\sin^2\theta(\cos\phi\left(2 a_6+2 a_{15}+c_6+c_{15}+d_{15}\right)/\sqrt{2}$&\multirow{2}{*}{-}&\multirow{2}{*}{-}\cr
&$- \sin\phi \left(a_6+a_{15}+b_6+b_{15}-d_6\right)\big)$&&\\\hline
\multirow{2}{*}{$\Xi^{0}_{c}\to n \eta^\prime $}      & $\sin^2\theta(\sin\phi\left(2 a_6+2 a_{15}+c_6+c_{15}+d_{15}\right)/\sqrt{2}$&\multirow{2}{*}{-}&\multirow{2}{*}{-}\cr
&$+\cos\phi  \left(a_6+a_{15}+b_6+b_{15}-d_6\right)\big)$&&\\\hline
\hline
\end{tabular}
\end{table*}

\end{widetext}

Note that for the SU(3) irreducible amplitudes in Eq.~\ref{su3} in fact each has two decay amplitudes. 
This means that the $a_i$, $b_i$,$c_i$, $d_i$ and $e_i$ each contains two amplitudes.
We expressed them by $f^q_i$ and $g^q_i$ representing parity violating scalar S and parity conserving pseudoscalar P wave form factors as the followings
\begin{eqnarray}
  q_6&=&G_F\bar{u}(f^q_6 - g^q_6\gamma_5)u,\quad q=a,b,c,d,\notag\\
  q_{15}&=&G_F\bar{u}(f^q_{15} - g^q_{15}\gamma_5)u,\quad q=a,b,c,d,e.
\end{eqnarray}
In general, $f^q$ and $g^q$ are complex. We find that the present data can be well fitted by all of them to be real. If CP violations are measured in some of the observables, we have to use complex numbers for the amplitudes. But for now, fitting the form factors with real numbers is sufficient. Their fitting results are displayed in Table~\ref{table3}.

The experimental observables, branching ratios, and polarization parameters $\alpha$ are given in Ref.~\cite{Huang:2021aqu} as the following
\begin{eqnarray}
\frac{d\Gamma}{d\cos\theta_M}&=&\frac{G_{F}^{2}|\vec{p}_{B_{n}}|(E_{B_n}+M_{B_{n}})}{8\pi M_{B_c}}(|F|^2+\kappa^2 |G|^2)\notag\\
&&\times(1+\alpha \hat\omega_i\cdot\hat p_{B_n} ),\label{decaywidth}
\end{eqnarray}
where $\hat\omega_i$ and $\hat p_{B_n}$ are the unit vector of initial state spin and final state momentum, respectively. Depending on the specific processes, the $F$ and $G$ are linear functions of $f_i$ and $g_i$, respectively.  The parameter $\alpha$~\cite{He:2015fsa} is given by
\begin{eqnarray}
\alpha&=&2\rm{Re}(F*G)\kappa/(|F|^2+\kappa^2 |G|^2), \notag\\
\kappa&=&|\Vec{p}_{B_n}|/(E_{B_n}+M_{B_n}).
\end{eqnarray}

As we discussed above, the parameters $a_6$ and $a_{15}$ can only be determined by the decays with $\eta$ or $\eta^\prime$ in the final states. Unfortunately, for the SU(3) irreducible amplitudes $a_6$ and $a_{15}$ only the combination $a_6 - a_{15}$ can be determined by existing date, but not $a_6 + a_{15}$.  For convenience, one redefines the new SU(3) irreducible amplitude and corresponding form factors as
\begin{eqnarray}
 a&=&a_6-a_{15},\quad a^\prime=a_6+a_{15},\notag\\
 f^a&=&f^a_6-f^a_{15}, \quad g^a=g^a_6-g^a_{15},\notag\\
 f^{a\prime}&=&f^a_6+f^a_{15}, \quad g^{a\prime}=g^a_6+g^a_{15}.
\end{eqnarray}
 
One can see that although there are 28 experimental data until now, the SU(3)  irreducible amplitude $a^\prime$ and its corresponding form factors $f^{a\prime},\;g^{a\prime}$ still cannot be determined. The 8 decays: $\Xi_c^0\to\Xi^0\eta^{(\prime)},\Xi_c^0\to\Sigma^0\eta^{(\prime)},\Xi_c^0\to\Lambda^0\eta^{(\prime)},\Xi_c^0\to n \eta^{(\prime)}$ rely on these form factors, but none of them have been measured. Therefore we eagerly waiting for data from future experimental facilities.
 
We performed a global fit with updated experimental data using the nonlinear least--$\chi^2$ method~\cite{Zhao:2021xwl} and present the fit results in Table~\ref{table3}.  The $\chi^2/d.o.f.$ is only 1.21 representing a good global fit. It is noteworthy that our fitted value of $\alpha(\Lambda_c\to\Sigma^0K^+)=-0.9958\pm0.0045$ has a $2\sigma$ standard deviation from the experimental data $\alpha(\Lambda_c\to\Sigma^0 K^+)_{exp}=-0.54\pm0.18\pm0.09$ , which was first measured by Belle in 2022 and contributed significantly to the $\chi^2$. This requires further verification with experimental data. We anticipate that future measurements from Belle and other experiments will help to clarify this issue.
 
By using the fitted form factors, we can predict other channels of anti-triplet charmed baryon hadronic two-body weak decays which have not yet been measured in experiments. Our predictions of these decays are presented in Tables~\ref{table1} and \ref{tableta}. These predictions can be used to test the validity of the SU(3) flavor symmetry.

\section{The $a'$ dependent branching ratios}\label{se_dis}

As we discussed in the above section, although we have 28 experimental data now, the SU(3)  irreducible amplitude $a^\prime$ and its corresponding form factors $f^{a\prime},\;g^{a\prime}$ still cannot be determined. This also results in the branching ratios of the 8 decay modes $\Xi_c^0\to\Xi^0\eta^{(\prime)},\Xi_c^0\to\Sigma^0\eta^{(\prime)},\Xi_c^0\to\Lambda^0\eta^{(\prime)},\Xi_c^0\to n \eta^{(\prime)}$ cannot be predicted.   In order to make predictions for these decays and guide experimental searches for them, we need to have some idea about the decay amplitude $a'$. Unfortunately, as previously mentioned, theoretical calculations for this parameter are not yet available. Thus, we turn the argument around to make predictions for them as a function of the $a'$ related form factors $f^{a\prime}$ and $g^{a\prime}$ to guide the experimental search for these decays. 

We now carry out an analysis by setting the two form factors $f^{a\prime},\;g^{a\prime}$ in a suitable region. To this end, we use two parameters $r^f$ and $r^g$ to show the ratio of the form factor corresponding to the SU(3) irreducible amplitude $a$ and  $a^\prime$, 
\begin{eqnarray}
r^f=f^{a\prime}/f^{a},\quad  r^g=g^{a\prime}/g^{a}.
\end{eqnarray}
Since the absolute values of $f^a$ and $g^a$ are expected to be similar in size to those of $f^{a'}$ and $g^{a'}$ , a reasonable range for $r^f$ and $r^g$ is between -1 and 1. In our plots that illustrate the dependence of various decay branching ratios, we will show a larger range to gain a better understanding of the trends of the dependence of these parameters.

The branching ratios of the undetermined 8 decays can be expressed as a function of the $r^f$ and $r^g$. 
Using the fit results in Table.~\ref{table3}, we obtain several interesting results. By setting one of the parameters $r^{f(g)}$ to be zero and  varying another parameter  $r^{g(f)}$, we give the curve of branching ratios which depend on the $r^{f(g)}$  with error band in Fig.~\ref{fig2}. In these curves, the dependence of the parameter $r^{f(g)}$ can be clearly seen. When the branching ratio varies slowly with changing the $r^{f(g)}$, we can give a rough prediction. If the branching ratio varies fast when we change the parameter $r^{f(g)}$, we can try to find the minimum and give the lower limit of the branching fraction.  

In Fig.~\ref{fig2}, we find that the branching ratios of decays which involve final state $\eta$ change slowly by varying both $r^f$ and $r^g$ respectively. In contrast, decays involving $\eta^\prime$ exhibit high dependence on $r^{f(g)}$. This can be explained by analyzing the SU(3) amplitude presented in Table~\ref{tableta}. Since the meson $\eta^{(\prime)}$ are the mixture of $\eta_q$ and $\eta_s$,the SU(3) amplitude of the decays involving these mesons should be expressed as a mixture of amplitudes with $\eta_q$ and $\eta_s$.

It can be observed that the parameter $a^\prime$ has opposite signs in the SU(3) amplitude for decay channels involving $\eta_q$ and $\eta_s$. The mixing angle $\phi=(39.3\pm1.0)^{\circ}$ causes the parameter $a^\prime$ from amplitudes with $\eta_q$ and $\eta_s$ to cancel out in amplitudes involving $\eta$, while they add up in amplitudes involving $\eta^\prime$. Therefore, the branching ratios of these decays with $\eta$ are less dependent on $r^{f(g)}$. Based on the above analysis, we can provide rough predictions of the branching ratios involving $\eta$ by varying  $r^{f(g)}\in[-1,1]$. The predictions for the ranges are
\begin{eqnarray}
&&Br(\Xi_c^0\to\Xi^0\eta) \sim[0.193,0.446]\%,\notag\\
&&Br(\Xi_c^0\to\Sigma^0\eta) \sim [0.0118,0.0333]\%,\notag\\
&&Br(\Xi_c^0\to\Lambda^0\eta) \sim [0.0039,0.0139]\%,\notag\\
&&Br(\Xi_c^0\to n\eta) \sim [0.00009, 0.00066]\%.\label{pred}
\end{eqnarray}
Although the branching ratios of decays with $\eta^\prime$ show very large $r^{f(g)}$ dependence in Fig.~\ref{fig2}, it also shows a very good  convergence around $r^{f(g)}=0$. Interestingly, we can give the lower limit of the branching ratio by finding its minimum value. 

For finding the minimum value of branching ratios in $r^{f}$-$r^g$ plane, we also show branching ratios of the 8  decays in Fig.~\ref{fig3}.
Fortunately, the branching ratios of all 8 decays show a very good convergence  and one can find its minimum value very easily.
\begin{widetext}
\begin{figure*}[htbp!]
  \begin{minipage}[t]{0.24\linewidth}
  \centering
\includegraphics[width=1\columnwidth]{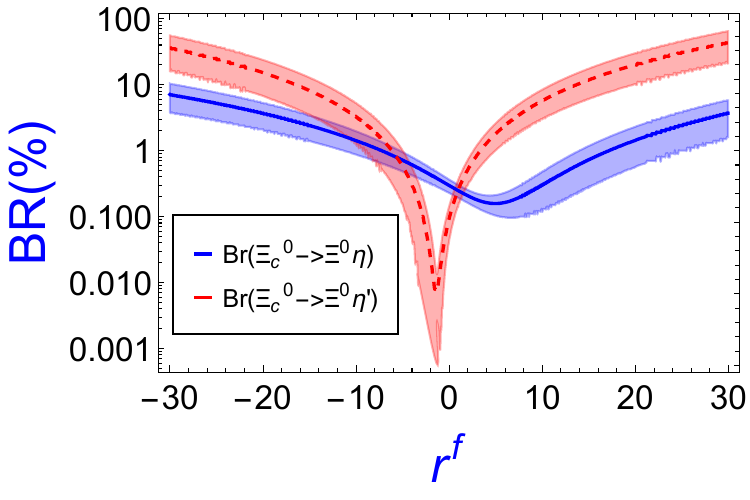} 
 \end{minipage}
   \begin{minipage}[t]{0.24\linewidth}
  \centering
\includegraphics[width=1\columnwidth]{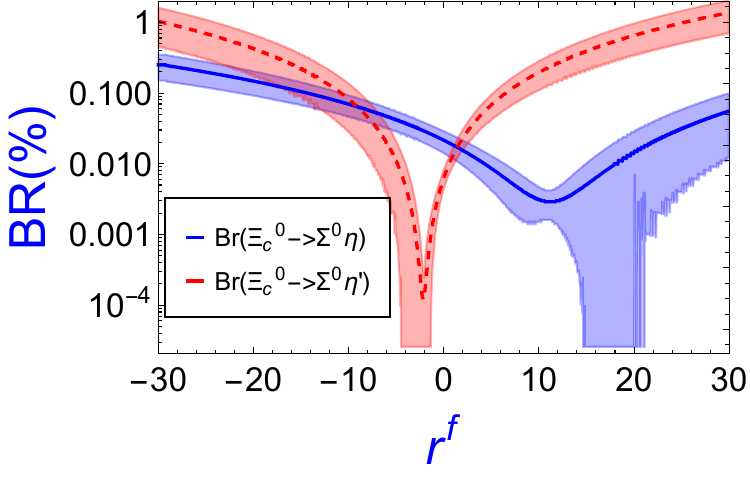} 
 \end{minipage}
   \begin{minipage}[t]{0.24\linewidth}
  \centering
\includegraphics[width=1\columnwidth]{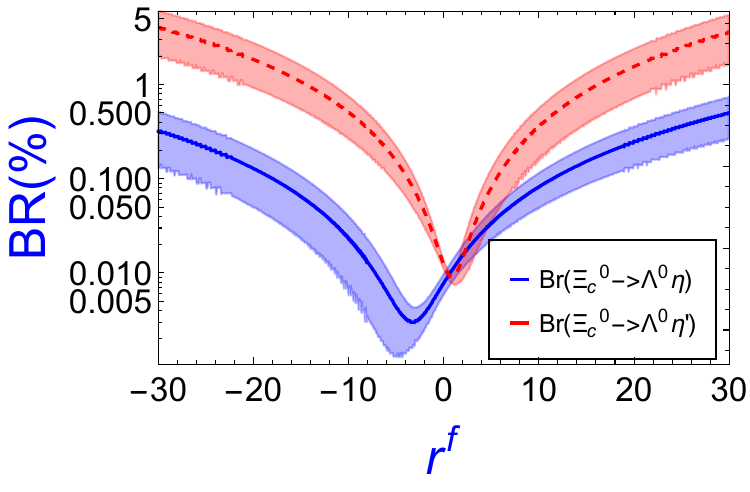} 
 \end{minipage}
   \begin{minipage}[t]{0.24\linewidth}
  \centering
\includegraphics[width=1\columnwidth]{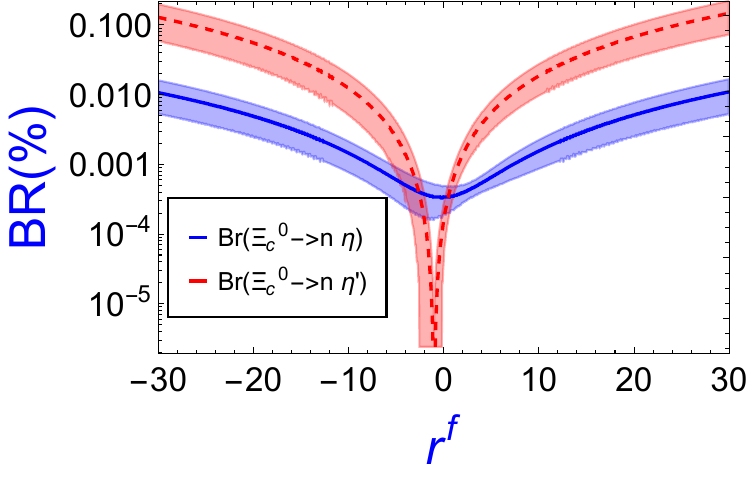} 
 \end{minipage}
  \begin{minipage}[t]{0.24\linewidth}
  \centering
\includegraphics[width=1\columnwidth]{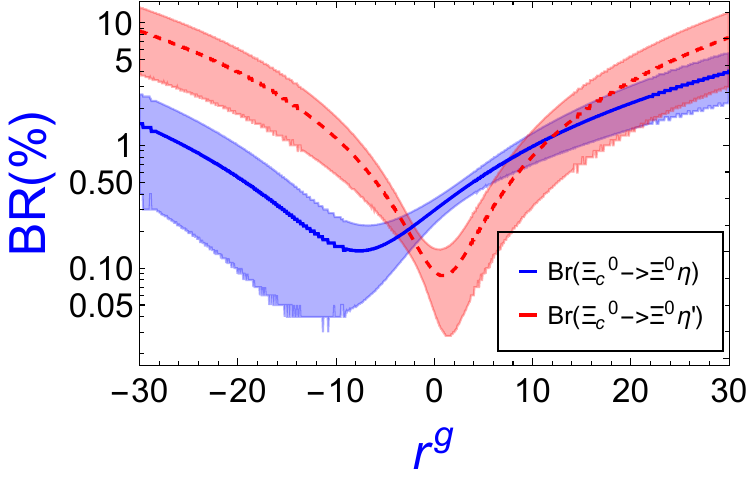} 
 \end{minipage}
   \begin{minipage}[t]{0.24\linewidth}
  \centering
\includegraphics[width=1\columnwidth]{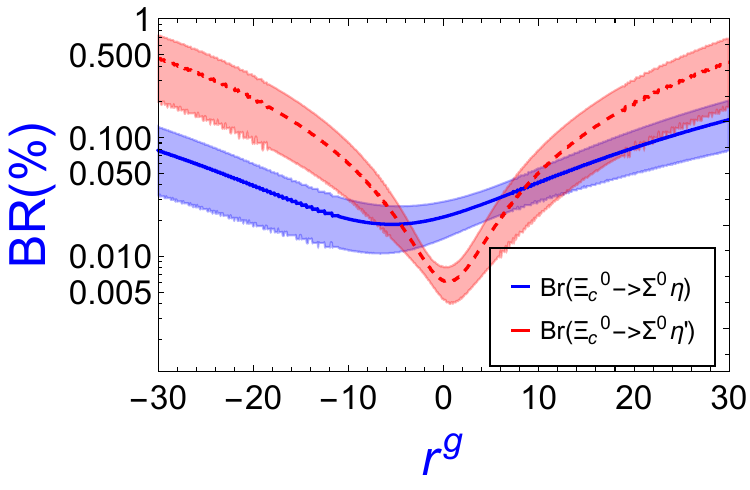} 
 \end{minipage}
   \begin{minipage}[t]{0.24\linewidth}
  \centering
\includegraphics[width=1\columnwidth]{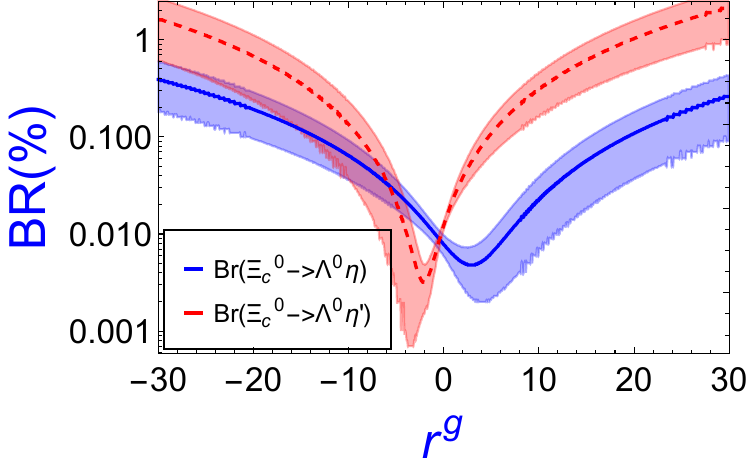} 
 \end{minipage}
   \begin{minipage}[t]{0.24\linewidth}
  \centering
\includegraphics[width=1\columnwidth]{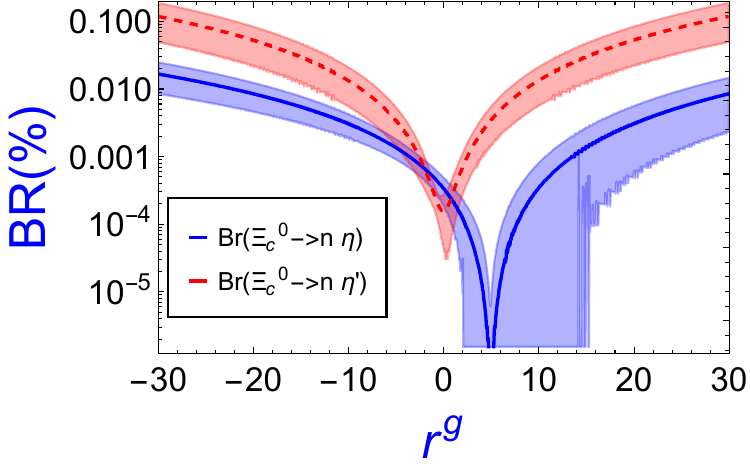} 
 \end{minipage}
 \caption{The branching ratios which depend on  the $r^f$ (first line) and $r^g$ (second line) for the 8 undetermined decays: $\Xi_c^0\to\Xi^0\eta^{(\prime)},\Xi_c^0\to\Sigma^0\eta^{(\prime)},\Xi_c^0\to\Lambda^0\eta^{(\prime)},\Xi_c^0\to n\eta^{(\prime)}$. In these figures, we set another parameter $r^{f(g)}=0$.}
\label{fig2}
\end{figure*}

\begin{figure*}[htbp!]
  \begin{minipage}[t]{0.24\linewidth}
  \centering
\includegraphics[width=1\columnwidth]{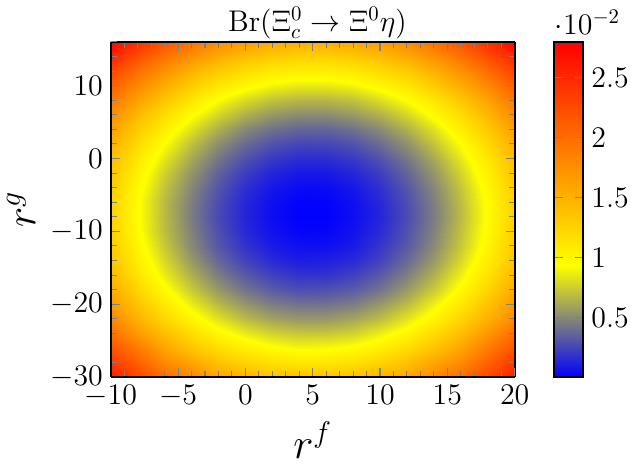} 
 \end{minipage}
   \begin{minipage}[t]{0.24\linewidth}
  \centering
\includegraphics[width=1\columnwidth]{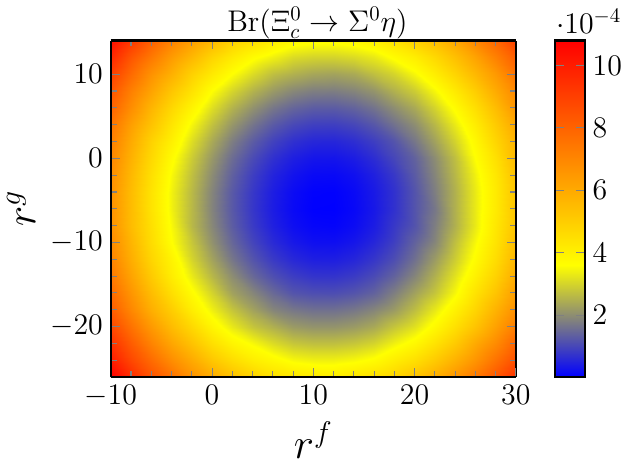} 
 \end{minipage}
   \begin{minipage}[t]{0.24\linewidth}
  \centering
\includegraphics[width=1\columnwidth]{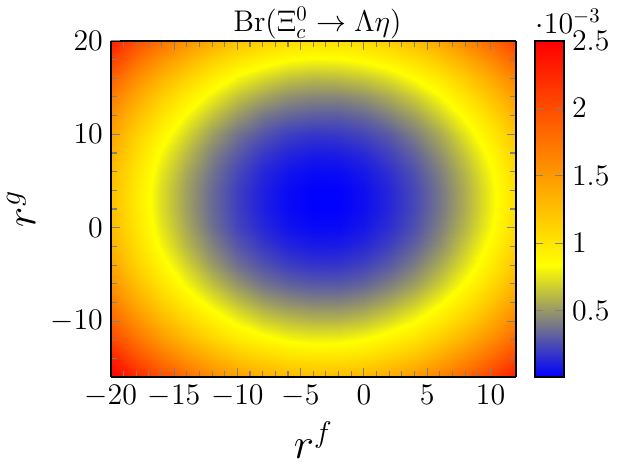} 
 \end{minipage}
   \begin{minipage}[t]{0.235\linewidth}
  \centering
\includegraphics[width=1\columnwidth]{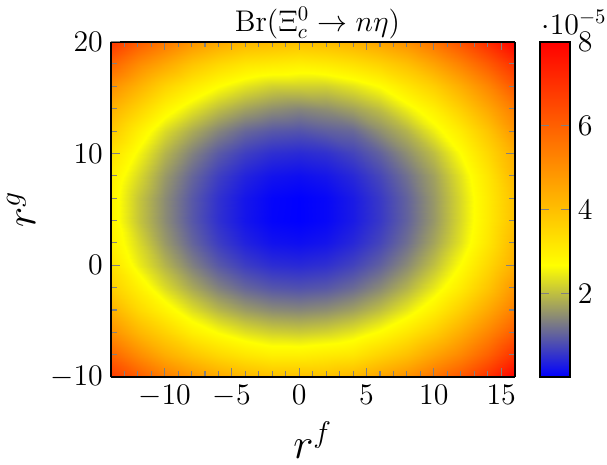} 
 \end{minipage}
   \begin{minipage}[t]{0.24\linewidth}
  \centering
\includegraphics[width=1\columnwidth]{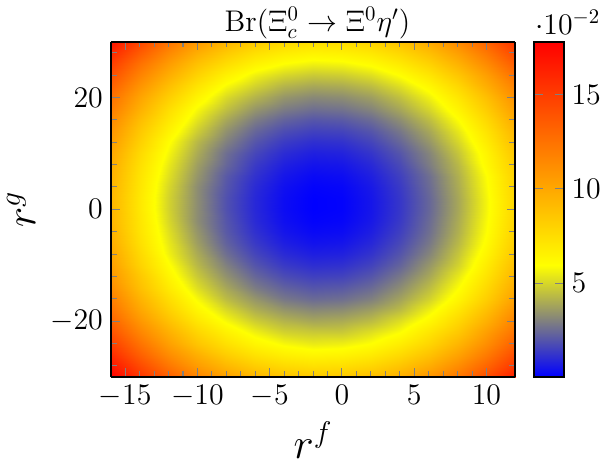} 
 \end{minipage}
   \begin{minipage}[t]{0.245\linewidth}
  \centering
\includegraphics[width=1\columnwidth]{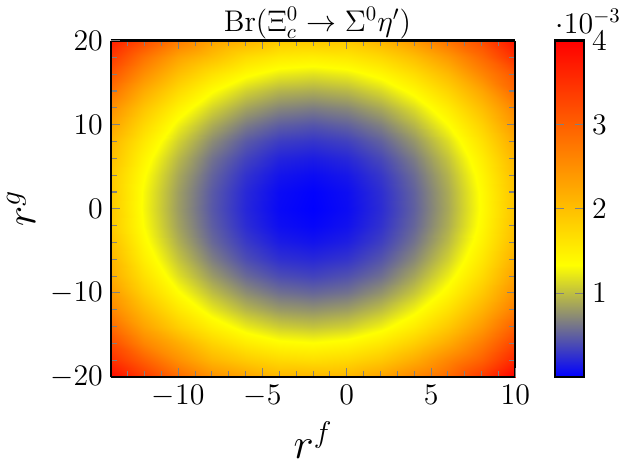} 
 \end{minipage}
   \begin{minipage}[t]{0.24\linewidth}
  \centering
\includegraphics[width=1\columnwidth]{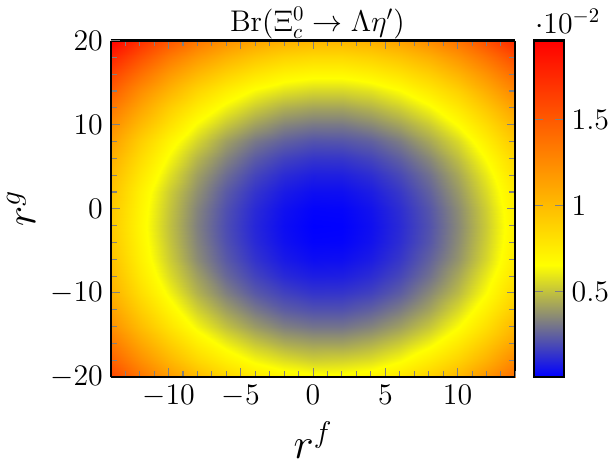} 
 \end{minipage}
   \begin{minipage}[t]{0.24\linewidth}
  \centering
\includegraphics[width=1\columnwidth]{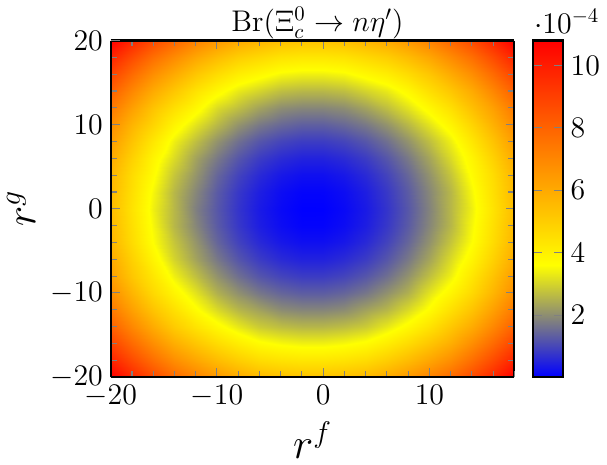} 
 \end{minipage}
 \caption{The branching ratios of the 8 undetermined decays: $\Xi_c^0\to\Xi^0\eta^{(\prime)},\Xi_c^0\to\Sigma^0\eta^{(\prime)},\Xi_c^0\to\Lambda^0\eta^{(\prime)},\Xi_c^0\to n\eta^{(\prime)}$ on  the $r^f-r^g$ plane.}
\label{fig3}
\end{figure*}
\end{widetext} 

For the branching ratios with $\eta^\prime$, we can find that its minimum value is around the $(r^f,r^g)=(0,0)$ which is consistent with the curve in Fig.~\ref{fig2}. After  scanning the branching fraction on  the $r^f-r^g$ plane, we give the lower limit as
\begin{eqnarray}
&&Br(\Xi_c^0\to\Xi^0\eta^\prime) \geq 0.002\%,\notag\\&&Br(\Xi_c^0\to\Sigma^0\eta^\prime) \geq 9\times 10^{-7},\notag\\
&&Br(\Xi_c^0\to\Lambda^0\eta^\prime) \geq 4.8\times 10^{-6},\notag\\&&Br(\Xi_c^0\to n\eta^\prime) \geq 6\times 10^{-8}.\label{predp}
\end{eqnarray}
It is evident that among these 8 decay channels, the branching ratio $Br(\Xi_c^0\to\Xi^0\eta)$ has the highest value and is at the 1\textperthousand ~level. This indicates that the probability of measuring the branching fraction $Br(\Xi_c^0\to\Xi^0\eta)$ experimentally is high.

\section{Conclusion}\label{se_sum}

A large amount of experimental data, including branching ratio and polarization asymmetry parameter $\alpha$, about anti-triplet charmed baryon hadronic two-body weak decays have been measured in Belle and BESIII.
Using the available 28 measured experimental data, we carried out an SU(3) symmetry analysis for these decays. 
We can obtain 16 SU(3) irreducible amplitudes with $\chi^2/d.o.f.$ of 1.21 indicating a very reasonable fit conforming to the results in Ref.~\cite{Huang:2021aqu} that SU(3) symmetry describes $T_{\bar 3} \to T_8 P$ very well.  This is a surprising one. The updated data and the fitting results are shown in Table.~\ref{table3}. In our global analysis, we noted that our prediction $\alpha(\Lambda_c\to\Sigma^0K^+)=-0.9958\pm0.0045$ which has $2\sigma$ standard deviation with experiment data $\alpha(\Lambda_c\to\Sigma^0 K^+)_{exp}=-0.54\pm0.18\pm0.09$.
Future measurements from experiments are expected to make further clarification of this observable.

Although we have a large number of experimental data, the two form factors $f^{a\prime}$ and $g^{a\prime}$ remain undetermined. That is because the two form factors can only be determined with 8 decays: $\Xi_c^0\to\Xi^0\eta^{(\prime)},\Xi_c^0\to\Sigma^0\eta^{(\prime)},\Xi_c^0\to\Lambda^0\eta^{(\prime)},\Xi_c^0\to n\eta^{(\prime)}$ in our SU(3) analysis, but none of them have been measured in the experiment. 
Nevertheless, by studying the dependence of these two form factors on these 8 decays, we can still provide rough predictions or lower limits on the branching ratios before the measurement in experimental facilities.
 The predictions or limits are given in Eq.~\ref{pred} and Eq.~\ref{predp}.  Our estimation will be very useful for experimental searches. Therefore, we eagerly await data from future experimental searches.
\section{Acknowledgements}
We thank Chengping Shen for pointing out a few typos in an earlier version. The work of X.G. He was supported by the Fundamental Research Funds for the Central Universities, by National Natural Science Foundation of P.R. China (No.12090064, 11735010, and 11985149) and by MOST 109–2112-M-002–017-MY3. The work of Z.P. Xing was supported by China Postdoctoral Science Foundation under Grant No. 2022M72210.

\appendix


\begin{thebibliography}{}
\bibitem{Belle:2019rba}
G.~Caria \textit{et al.} [Belle],
Phys. Rev. Lett. \textbf{124}, no.16, 161803 (2020)
doi:10.1103/PhysRevLett.124.161803
[arXiv:1910.05864 [hep-ex]].

\bibitem{Belle:2019oag}
A.~Abdesselam \textit{et al.} [Belle],
Phys. Rev. Lett. \textbf{126}, no.16, 161801 (2021)
doi:10.1103/PhysRevLett.126.161801
[arXiv:1904.02440 [hep-ex]].

\bibitem{LHCb:2021trn}
R.~Aaij \textit{et al.} [LHCb],
Nature Phys. \textbf{18}, no.3, 277-282 (2022)
doi:10.1038/s41567-021-01478-8
[arXiv:2103.11769 [hep-ex]].

\bibitem{LHCb:2022piu}
R.~Aaij \textit{et al.} [LHCb],
Phys. Rev. Lett. \textbf{128}, no.19, 191803 (2022)
doi:10.1103/PhysRevLett.128.191803
[arXiv:2201.03497 [hep-ex]].

\bibitem{Choudhury:2022ktg}
S.~Choudhury [Belle],
Springer Proc. Phys. \textbf{277}, 231-235 (2022)
doi:10.1007/978-981-19-2354-8\_42

\bibitem{LHCb:2023zxo}
 [LHCb],
[arXiv:2302.02886 [hep-ex]].

\bibitem{LHCb:2023cjr}
R.~Aaij \textit{et al.} [LHCb],
[arXiv:2305.01463 [hep-ex]].

\bibitem{Zhao:2018zcb}
Z.~X.~Zhao,
Chin. Phys. C \textbf{42}, no.9, 093101 (2018)
doi:10.1088/1674-1137/42/9/093101
[arXiv:1803.02292 [hep-ph]].

\bibitem{Geng:2019xbo}
C.~Q.~Geng, C.~W.~Liu and T.~H.~Tsai,
Phys. Lett. B \textbf{794}, 19-28 (2019)
doi:10.1016/j.physletb.2019.05.024
[arXiv:1902.06189 [hep-ph]].

\bibitem{Zou:2019kzq}
J.~Zou, F.~Xu, G.~Meng and H.~Y.~Cheng,
Phys. Rev. D \textbf{101}, no.1, 014011 (2020)
doi:10.1103/PhysRevD.101.014011
[arXiv:1910.13626 [hep-ph]].

\bibitem{Jia:2019zxi}
C.~P.~Jia, D.~Wang and F.~S.~Yu,
Nucl. Phys. B \textbf{956}, 115048 (2020)
doi:10.1016/j.nuclphysb.2020.115048
[arXiv:1910.00876 [hep-ph]].


\bibitem{Zhao:2018mov}
H.~J.~Zhao, Y.~L.~Wang, Y.~K.~Hsiao and Y.~Yu,
JHEP \textbf{02}, 165 (2020)
doi:10.1007/JHEP02(2020)165
[arXiv:1811.07265 [hep-ph]].

\bibitem{Xu:2021mkg}
F.~Xu, Q.~Wen and H.~Zhong,
LHEP \textbf{2021}, 218 (2021)
doi:10.31526/lhep.2021.218

\bibitem{Huang:2021aqu}
F.~Huang, Z.~P.~Xing and X.~G.~He,
JHEP \textbf{03}, 143 (2022)
[erratum: JHEP \textbf{09}, 087 (2022)]
doi:10.1007/JHEP03(2022)143
[arXiv:2112.10556 [hep-ph]].

\bibitem{Zhao:2021sje}
Z.~X.~Zhao,
[arXiv:2103.09436 [hep-ph]].

\bibitem{Groote:2021pxt}
S.~Groote and J.~G.~K\"orner,
Eur. Phys. J. C \textbf{82}, no.4, 297 (2022)
doi:10.1140/epjc/s10052-022-10224-0
[arXiv:2112.14599 [hep-ph]].

\bibitem{Li:2021iwf}
H.~B.~Li and X.~R.~Lyu,
Natl. Sci. Rev. \textbf{8}, no.11, nwab181 (2021)
doi:10.1093/nsr/nwab181
[arXiv:2103.00908 [hep-ex]].
\bibitem{Hsiao:2021nsc}
Y.~K.~Hsiao, Y.~L.~Wang and H.~J.~Zhao,
JHEP \textbf{09}, 035 (2022)
doi:10.1007/JHEP09(2022)035
[arXiv:2111.04124 [hep-ph]].


\bibitem{Zhong:2022exp}
H.~Zhong, F.~Xu, Q.~Wen and Y.~Gu,
JHEP \textbf{02}, 235 (2023)
doi:10.1007/JHEP02(2023)235
[arXiv:2210.12728 [hep-ph]].

\bibitem{Zhao:2023yuk}
Z.~X.~Zhao, F.~W.~Zhang, X.~H.~Hu and Y.~J.~Shi,
[arXiv:2304.07698 [hep-ph]].

\bibitem{Liu:2023pyk}
H.~Liu and C.~Yang,
[arXiv:2304.12128 [hep-ph]].

\bibitem{Wang:2023don}
D.~Wang,
[arXiv:2301.07443 [hep-ph]].

\bibitem{BESIII:2017fim}
M.~Ablikim \textit{et al.} [BESIII],
Phys. Rev. D \textbf{95}, no.11, 111102 (2017)
doi:10.1103/PhysRevD.95.111102
[arXiv:1702.05279 [hep-ex]].

\bibitem{BESIII:2018cvs}
M.~Ablikim \textit{et al.} [BESIII],
Phys. Lett. B \textbf{783}, 200-206 (2018)
doi:10.1016/j.physletb.2018.06.046
[arXiv:1803.04299 [hep-ex]].

\bibitem{BESIII:2018cdl}
M.~Ablikim \textit{et al.} [BESIII],
Chin. Phys. C \textbf{43}, no.8, 083002 (2019)
doi:10.1088/1674-1137/43/8/083002
[arXiv:1811.08028 [hep-ex]].

\bibitem{Belle:2021crz}
Y.~B.~Li \textit{et al.} [Belle],
Phys. Rev. Lett. \textbf{127}, no.12, 121803 (2021)
doi:10.1103/PhysRevLett.127.121803
[arXiv:2103.06496 [hep-ex]].

\bibitem{Belle:2021mvw}
S.~X.~Li \textit{et al.} [Belle],
Phys. Rev. D \textbf{103}, no.7, 072004 (2021)
doi:10.1103/PhysRevD.103.072004
[arXiv:2102.12226 [hep-ex]].

\bibitem{Li:2021bff}
L.~Li [Belle],
PoS \textbf{ICHEP2020}, 391 (2021)
doi:10.22323/1.390.0391
[arXiv:2102.03703 [hep-ex]].

\bibitem{Lyu:2021biq}
X.~R.~Lyu [BESIII],
PoS \textbf{CHARM2020}, 016 (2021)
doi:10.22323/1.385.0016


\bibitem{BESIII:2022izy}
M.~Ablikim \textit{et al.} [BESIII],
Phys. Rev. D \textbf{106}, no.7, 072002 (2022)
doi:10.1103/PhysRevD.106.072002
[arXiv:2207.14461 [hep-ex]].

\bibitem{BESIII:2022tnm}
M.~Ablikim \textit{et al.} [BESIII],
Phys. Rev. D \textbf{106}, no.11, L111101 (2022)
doi:10.1103/PhysRevD.106.L111101
[arXiv:2208.04001 [hep-ex]].

\bibitem{BESIII:2022wxj}
M.~Ablikim \textit{et al.} [BESIII],
Phys. Rev. D \textbf{106}, no.5, 052003 (2022)
doi:10.1103/PhysRevD.106.052003
[arXiv:2207.10906 [hep-ex]].

\bibitem{BESIII:2022bkj}
M.~Ablikim \textit{et al.} [BESIII],
Phys. Rev. Lett. \textbf{128}, no.14, 142001 (2022)
doi:10.1103/PhysRevLett.128.142001
[arXiv:2201.02056 [hep-ex]].

\bibitem{Belle:2021vyq}
S.~X.~Li \textit{et al.} [Belle],
JHEP \textbf{03}, 090 (2022)
doi:10.1007/JHEP03(2022)090
[arXiv:2112.14276 [hep-ex]].

\bibitem{Belle:2022uod}
L.~K.~Li \textit{et al.} [Belle],
Sci. Bull. \textbf{68}, 583-592 (2023)
doi:10.1016/j.scib.2023.02.017
[arXiv:2208.08695 [hep-ex]].

\bibitem{Belle:2022bsi}
S.~X.~Li \textit{et al.} [Belle],
Phys. Rev. D \textbf{107}, 032003 (2023)
doi:10.1103/PhysRevD.107.032003
[arXiv:2208.10825 [hep-ex]].

\bibitem{He:2018joe}
X.~G.~He, Y.~J.~Shi and W.~Wang,
Eur. Phys. J. C \textbf{80}, no.5, 359 (2020)
doi:10.1140/epjc/s10052-020-7862-5
[arXiv:1811.03480 [hep-ph]].

\bibitem{Gan:2020aco}
L.~Gan, B.~Kubis, E.~Passemar and S.~Tulin,
Phys. Rept. \textbf{945}, 1-105 (2022)
doi:10.1016/j.physrep.2021.11.001
[arXiv:2007.00664 [hep-ph]].

\bibitem{ParticleDataGroup:2022pth}
R.~L.~Workman \textit{et al.} [Particle Data Group],
PTEP \textbf{2022}, 083C01 (2022)
doi:10.1093/ptep/ptac097

\bibitem{He:2015fsa}
M.~He, X.~G.~He and G.~N.~Li,
Phys. Rev. D \textbf{92}, no.3, 036010 (2015)
doi:10.1103/PhysRevD.92.036010
[arXiv:1507.07990 [hep-ph]].

\bibitem{Zhao:2021xwl}
P.~.Lepage and C.~ Gohlke,
gplepage/lsqfit:~lsqfit~version~ 11.7~(v11.7),Zenodo~doi.org/10.5281/zenodo.4037174

\end{thebibliography}
\end{document}